# Interaction of Bianchi Type-I Anisotropic Cloud String Cosmological Model Universe with Electromagnetic Field

Kangujam Priyokumar Singh[1], Mahbubur Rahman Mollah[2], Rajshekhar Roy Baruah[3] and Meher Daimary[4, *]

[1,3,4]Department of Mathematical Sciences, Bodoland University, Kokrajhar, Kokrajhar-783370, BTC, Assam, India

[2]Department of Mathematics, Commerce College, Kokrajhar, Kokrajhar-783370, BTC, Assam, India

*Corresponding author: meherdaimary@gmail.com

**Abstract:** Here, we have investigated the interaction of Bianchi type-I anisotropic cloud string cosmological model universe with electromagnetic field in the context of general relativity. In this paper, the energy momentum tensor is assumed to be the sum of the rest energy density and string tension density with an electromagnetic field. To obtain exact solutions of Einstein's field equations, we take average scale factor as an integrating function of time. Also, the dynamics and significance of various physical parameters of model are discussed.

## 1. Introduction

Various kinds of literature reveal that our universe is expanding with an accelerated rate. In FRW space-time, Einstein's theory of gravitation is considered to be a perfect theory to explain our expanding universe. Beyond the standard model ΛCDM in extending cosmology, Bull et al. [1] studied the alternative cosmology and summarize the current status of ΛCDM as a physical theory. On a large scale, FRW space-time describes only isotropic and homogeneous universe. But recent observations and arguments suggest that there exist an anisotropic phase before reaching to an isotropic one during the cosmic expansion of the universe. Bianchi type cosmological models represent the homogeneous and anisotropic universe, where the isotropy nature of them may also be studied with the passage of time. Also, in the theoretical perspective, the anisotropic universes have possessed a greater generality than the isotropic model universes. Many authors [2, 3, 4, 5, 6, 7] investigated anisotropic Bianchi type cosmological model in different perspectives. Investigating Bianchi type metric, Zel'dovich [8] obtained the general solution of the Einstein's field equations for dust case and the singularity behaviour of the solution was also explained by him. Kibble [9], in his paper "Topology of Cosmic Domains and Strings", recognized the stable topological defects that occurred throughout the phase transition as strings. Also, he showed that the topological defects of domain structure depends on the homogeneity group of the manifold of degenerate vacua. In order to study String Cosmology, Letelier [10] choose Bianchi type I and Kantowski-Sachs space-time cosmological model. Collins and Hawking [11] showed that the Bianchi type VII models are the most general uniform cosmological models which are infinite in all three spatial directions. These models have wide class initial conditions providing the maximum possible number of arbitrary constants. He also showed that all the initial anisotropic universes do not approach isotropy but the subclass of the universe having escape velocity approach isotropy.

Due to the key role of string in describing the evolution of early stage of our universe, many authors in recent times, extensively studied the string cosmological models of the universe. The string can describe simultaneously the nature and fundamental configuration of the early universe. String theory is the most actively studied approach to quantum gravity, and by using this theory we can discuss the physics of early universe. String theory provides us with a single theoretical structure, where all the matters and forces are united and describes the early stage of our universe in terms of vibrating strings rather than particles. According to Kibble [9], cosmic strings are nothing but stable line like topological objects/defects which are formed at some stage in the phase transition in premature early period of our universe. According to GUT (grand unified theories) (Zel'dovich et al., [12] Kibble, [9, 13] Everett, [14] Vilenkin [15, 16]) there is a symmetry breaking during the phase transition in the early stage of universe after the big-bang and these strings arise when the cosmic temperature goes down below some critical temperature. Therefore, in order to study about the early stage of universe, strings can play a crucial role. Also, large scale structures like galaxies, and cluster of galaxies are formed by massive closed loops of strings. The cosmic strings couple with the gravitational field and may possess stress-energy. Therefore, the study of the gravitational effects arising from the cosmic strings are treated as one of the interesting work.

Letelier revolutionized the general theory of relativity by introducing strings in it. In 1979, Letelier [17] obtained the general solution of Einstein's field equations for a cloud of strings with spherical, plane and a particular case of cylindrical symmetry. Also, in 1983, Letelier [10] obtained massive string cosmological models in Bianchi type-I and Kantowski-Sachs space-times. After Letelier, many authors studied string cosmological models in different circumstances.

Considering Bianchi type-II, -VI0, -VIII and -IX space-times, Krori [18] and Wang [19] studied the Letelier string cosmological model and obtained their exact solutions. In the Riemann-Cartan space-time, Smalley [20] studied string fluid dynamical model within the framework of Einstein-Cartan theory. Investigating LRS Bianchi type-I metric, Xing, [21] obtained an exact solution string cosmological model with bulk viscosity by assuming the coefficient of bulk viscosity as a power function of energy density. In spherical symmetric space-time, Yavuz [22] examined charged strange quark matter attached to the string cloud and showed that one-parameter group of conformal motions is exist. Yilmaz [23] obtained the Kaluza-Klein cosmological solutions for quark matter attached to the string cloud in the context of general relativity. Investigating a Bianchi type-V space-time in a scalar-tensor theory, Rao [24] obtained an exact perfect fluid cosmological model based on Lyra manifold, when the displacement vector $\beta$ is a constant. But if $\beta$ is a function of cosmic time $t$ then only in case of radiation this model exists. Tripathy [25] studied an anisotropic and spatially homogeneous Bianchi type-VI0 space-time and obtained string cloud cosmological models in Saez-Ballester Scalar-Tensor theory. Adhav et al. [26] obtained string cosmological models in Brans-Dicke theory of gravitation, where they had solved the field equations by using the condition that the sum of the tension density and energy density is zero. Also, Pawar [27] studied Kaluza Klein string cosmological model in the framework of f(R, T) theory of gravity and solved the field equations by assuming a power law relation between scale factor and a time varying deceleration parameter.

The cosmological model in presence of electromagnetic field plays an important role in the evolution of universe and the formation of large scale structures like galaxies and other stellar bodies. The present phase of accelerated expansion of the universe is due to the presence of a cosmological electromagnetic field generated during inflation. Jimenez and Maroto [28] showed that the presence of a temporal electromagnetic field on cosmological scales generates an effective cosmological constant which is accounted for accelerated expansion of the universe. Tripathy et al. [29] have investigated an inhomogeneous string cosmological model with electromagnetic field in general relativity. Recently, Parikh [30] investigated a Bianchi type II string dust cosmological model with an electromagnetic field in

Lyra's Geometry. Also, in presence of magnetic field in f (R, T) theory of gravity, Pradhan [31] studied a class of anisotropic and homogeneous Bianchi type-V cosmological models with massive strings. Grasso [32] analyzed a large variety of aspects of magnetic fields in the early Universe and discussed about recent time fields and their evolution in galaxies and clusters of galaxies. The breakdown of statistical homogeneity and isotropy is also due to the magnetic field. The large scale galaxies and clusters of galaxy host magnetic fields. Subramanian [33] on his paper showed that primordial magnetic fields have greatly influence on formation of stellar structures, especially on dwarf galaxy scales.

Motivated from the study of above literatures, in this paper, we have solved the Einstein's field equations for energy momentum tensor for cloud string interacting with electromagnetic field and discussed in detail all aspects of physical and kinematic properties. The paper is organized as follows: In Section 2, the metric and the field equations are presented; In Section 3, we deal with solution of the field equations by assuming average scale factor as $a = (t^k e^t)^{\frac{1}{l}}$; Physical properties of our model universe are presented in Section 4; Section 5 deals with illustrations and figures. Finally, in Section 6, we summarize the results and give concluding remarks.

## 2. The Metric and Field Equations

We consider the spatially homogeneous and anisotropic Bianchi type-I metric in the form

$$ds^2 = -dt^2 + A^2 dx^2 + B^2 dy^2 + C^2 dz^2, \qquad (1)$$

where $A, B$ and $C$ are functions of cosmic time $t$ only.

The energy-momentum tensor for a cloud string has the form

$$T_{ij} = \rho u_i u_j - \lambda x_i x_j + E_{ij}, \qquad (2)$$

Where, $\rho$ being the rest energy density of the cloud of strings with particles attached to them, $\rho = \rho_p + \lambda$, $\rho_p$ being the particle density and $\lambda$ the tension density of the string. $x_i$ is a unit space-like vector representing the direction of strings so that $x^2 = 0 = x^3 = x^4$ and $x^1 \neq 0$, and $u_i$ is the four velocity vector satisfying the following conditions

$$u^i u_i = -x^i x_i = -1, \qquad (3)$$

and

$$u^i x_i = 0, \qquad (4)$$

The four velocity vector $u^i$ and the direction of string $x^i$ are given by

$$u^i = (0,0,0,1) \tag{5}$$

and

$$x^i = (\tfrac{1}{A}, 0, 0, 1) \tag{6}$$

where the direction of strings are parallel to $x$-axis.

The electromagnetic field $E_{ij}$ which is a part of the energy momentum tensor is considered as

$$E_{ij} = \tfrac{1}{4\pi}(g^{\alpha\beta} F_{i\alpha} F_{i\beta} - \tfrac{1}{4} g_{ij} F_{\alpha\beta} F^{\alpha\beta}) \tag{7}$$

where $F_{\alpha\beta}$ is the electromagnetic field tensor.

If we quantize the magnetic field along the $x$ axis then $F_{14}$ is the only non-vanishing component of the electromagnetic field tensor $F_{ij}$. Therefore, if we assume infinite electromagnetic conductivity, it can be shown that $F_{12} = F_{13} = F_{23} = F_{24} = F_{34} = 0$.

Therefore, from equation (7), the nontrivial components of the electromagnetic field $E_{ij}$ are obtained as follows

$$E_1^1 = -E_2^2 = -E^3 = E_1^1 = -\tfrac{1}{2} g^{11} g^{44} F_{14}^2 = \tfrac{1}{2A^2} F_{14}^2, \tag{8}$$

$$OR$$

In co-moving coordinates, if we take the magnetic field along $x$- axis then $F_{23}$ is the only non-vanishing component of the electromagnetic field tensor $F_{ij}$. Therefore, if we assume infinite electromagnetic conductivity, it can be shown that $F_{14} = F_{24} = F_{34} = F_{12} = F_{13} = 0$

Therefore, from equation (7), the nontrivial components of the electromagnetic field $E_{ij}$ are obtained as follows

$$E_1^1 = -E_2^2 = -E^3 = E_1^1 = \tfrac{1}{2A^2} F_{23}^2$$

The Einstein's field equations (with $\tfrac{8\pi G}{c^4} = 1$) is given by

$$R_i^j - \tfrac{1}{2} R g_i^j = -T_i^j \tag{9}$$

where $R_i^j$ is the Ricci tensor, $R = g^{ij} R_{ij}$ is the Ricci scalar.

The important physical quantities like the spatial volume $V$, the average scale factor $R$, the expansion scalar $\theta$, the Hubble expansion factor $H$, deceleration parameter $q$, the shear scalar $\sigma^2$ and the mean anisotropy parameter $\Delta$ for the metric (1) are found as follows:

$$V = R^3 = ABC, \tag{10}$$

$$\theta = u^i_{;i} = \frac{\dot{A}}{A} + \frac{\dot{B}}{B} + \frac{\dot{C}}{C} \tag{11}$$

$$H = \frac{\dot{R}}{R} = \frac{1}{3}\left(\frac{\dot{A}}{A} + \frac{\dot{B}}{B} + \frac{\dot{C}}{C}\right) \tag{12}$$

$$q = -\frac{R\ddot{R}}{\dot{R}^2} = \frac{d}{dt}\left(\frac{1}{H}\right) - 1 \tag{13}$$

$$\sigma^2 = \frac{1}{2}\sigma_{ij}\sigma^{ij} = \frac{1}{2}\left(\frac{\dot{A}^2}{A^2} + \frac{\dot{B}^2}{B^2} + \frac{\dot{C}^2}{C^2}\right) - \frac{1}{6}\theta^2 \tag{14}$$

$$\Delta = \frac{1}{3}\sum_{i=1}^{3}\left(\frac{H_i - H}{H}\right)^2 \tag{15}$$

where an over head dot stands for the first and double over head dot for the second derivative with respect to cosmic time $t$.

The Einstein's field equations (9) together with (2) for the line-element (1) reduces to the following system of equations:

$$\frac{\ddot{B}}{B} + \frac{\ddot{C}}{C} + \frac{\dot{B}\dot{C}}{BC} = \lambda - \frac{1}{2A^2}F_{14}^2 \tag{16}$$

$$\frac{\ddot{A}}{A} + \frac{\ddot{C}}{C} + \frac{\dot{A}\dot{C}}{AC} = \frac{1}{2A^2}F_{14}^2 \tag{17}$$

$$\frac{\ddot{A}}{A} + \frac{\ddot{B}}{B} + \frac{\dot{A}\dot{B}}{AB} = \frac{1}{2A^2}F_{14}^2 \tag{18}$$

$$\frac{\dot{A}\dot{B}}{AB} + \frac{\dot{B}\dot{C}}{BC} + \frac{\dot{A}\dot{C}}{AC} = \rho - \frac{1}{2A^2}F_{14}^2 \tag{19}$$

### 3. Solution of the Field Equations

The field equations $(16) - (19)$ are a system of four equations in five unknown parameters $A, B, C, \rho, \lambda$ and $F_{14}$. Therefore, two additional constraint relating these parameters is required to obtain explicit solutions of the system. These two relations are taken as

(i) The shear scalar $\sigma$ is proportional to the scalar expansion $\theta$ (Collins et. al. [34], VUM Rao et. al. [35]) so that we may take

$$B = C^n, \tag{20}$$

where $n$ is constant, and

(ii) the Hubble parameter H is related to the average scale factor R by the relation as taken by Berman [36] and Ram et al. [37]. By using this type of relation, Berman [36], Berman and Gomide [38] solved FRW models whereas Ram et al. [37] solved Bianchi Type V cosmological models in Lyra's Geometry which is considered as

$$H = aR^{-m} \tag{21}$$

where $a > 0$ and $m \geq 0$ are constants.

Equation (21) gives us

$$R = (amt + b)^{\frac{1}{m}} \qquad if\ m \neq 0 \qquad (22)$$

Equation (21) gives us

$$R = ce^{at} \qquad if\ m = 0 \qquad (23)$$

where $b$ and $c$ are constants.

Therefore, from equation (13), the deceleration parameter $q$ is obtained that

$$q = m - 1 \qquad if\ m \neq 0 \qquad (24)$$

and

$$q = -1 \qquad if\ m = 0 \qquad (25)$$

**Case-I: When $m \neq 0$**

The equations (17), (18), (20) and (22) will give us

$$C = k_2^{\frac{1}{n-1}} e^{\frac{k_1(amt+b)^{\frac{m-3}{m}}}{a(m-3)(n-1)}} \qquad (26)$$

Where $k_1, k_2$ are constants. Without loss of generality, we may assume $k_1 = k_2 = 0$ so that we have

$$C = e^{\frac{(amt+b)^{\frac{m-3}{m}}}{a(m-3)(n-1)}} \qquad (27)$$

Therefore, the equations (10), (20) and (22) will give us

$$A = (amt + b)^{\frac{3}{m}} e^{-\frac{(n+1)(amt+b)^{\frac{m-3}{m}}}{a(m-3)(n-1)}} \qquad (28)$$

and

$$B = e^{\frac{n(amt+b)^{\frac{m-3}{m}}}{a(m-3)(n-1)}} \qquad (29)$$

Therefore, the line element (1) becomes

$$ds^2 = -dt^2 + (amt+b)^{\frac{6}{m}} e^{-\frac{2(n+1)(amt+b)^{\frac{m-3}{m}}}{a(m-3)(n-1)}} dx^2 + e^{\frac{2n(amt+b)^{\frac{m-3}{m}}}{a(m-3)(n-1)}} dy^2 +$$

$$e^{\frac{2(amt+b)^{\frac{m-3}{m}}}{a(m-3)(n-1)}} dz^2 \qquad (30)$$

Equation (30) represents a Bianchi Type I cosmological model universe with electromagnetic field with special law of Hubble parameter.

## 4. Some Physical Properties of the Model

Now, for our model universe given by equation (30), the values of energy density $\rho$, String tension density $\lambda$ and electromagnetic field tensor $F_{14}$ can be obtained from equations (16) - (19), as

$$\rho = -3a^2(m-3)(amt+b)^{-2} \tag{31}$$

$$\lambda = -3a^2(m-3)(amt+b)^{-2} - \frac{6a(n+1)}{n-1}(amt+b)^{-\frac{m+3}{m}} + \frac{2(n^2+n+1)}{(n-1)^2}(amt+b)^{-\frac{6}{m}} \tag{32}$$

$$F_{14} = \sqrt{2}\left[-3a^2(m-3)(amt+b)^{-\frac{2(m-3)}{m}} - \frac{3a(n+1)}{n-1}(amt+b)^{-\frac{m-3}{m}} + \frac{n^2+n+1}{(n-1)^2}\right]^{\frac{1}{2}} e^{\frac{(n+1)(amt+b)^{\frac{m-3}{m}}}{a(m-3)(n-1)}} \tag{33}$$

Now, by using the relation $\rho = \rho_p + \lambda$ we have

$$\rho_p = \frac{6a(n+1)}{n-1}(amt+b)^{-\frac{m+3}{m}} - \frac{2(n^2+n+1)}{(n-1)^2}(amt+b)^{-\frac{6}{m}} \tag{34}$$

Again, the physical quantities such as proper volume $V$, Hubble parameter $H$, expansion scalar $\theta$, shear scalar $\sigma^2$, mean anisotropy parameter $\Delta$ are obtained as follows:

The proper volume $V$ is given by

$$V = (amt+b)^{\frac{3}{m}} \tag{35}$$

The Hubble parameter $H$ is in the form

$$H = \frac{a}{amt+b} \tag{36}$$

The expansion scalar $\theta$ is

$$\theta = \frac{3a}{amt+b} \tag{37}$$

The shear scalar $\sigma^2$ is

$$\sigma^2 = 3a^2(amt+b)^{-2} - \frac{3a(n+1)}{n-1}(amt+b)^{-\frac{m+3}{m}} + \frac{n^2+n+1}{(n-1)^2}(amt+b)^{-\frac{6}{m}} \tag{38}$$

The mean anisotropy parameter $\Delta$ is given by

$$\Delta = 2 - \frac{2(m+1)}{a(n-1)}(amt+b)^{\frac{m-3}{m}} + \frac{2(n^2+n+1)}{3a^2(n-1)^2}(amt+b)^{\frac{2(m-3)}{m}} \tag{39}$$

From (37) and (38), we have

$$\lim_{t \to \infty} \frac{\sigma^2}{\theta^2} = \frac{(m-1)^2}{3(m+2)^2} = Constant (\neq 0 \text{ for } m \neq 1) \tag{40}$$

Taking $a = 1$, $b = 0$, $m = n = 0.5$, the variation of some of the parameters are shown in the next page.

# 5. Illustrations and Figures

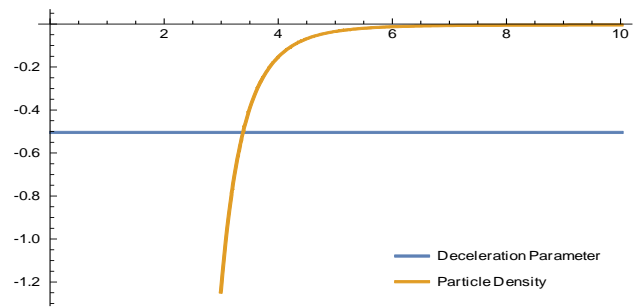

Figure 1: Variation of $q$, $\rho_p$ vs. Time $t$ in Gyr.

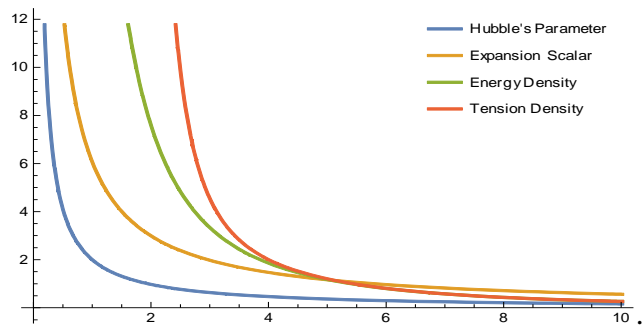

Figure 2: Variation of $H$, $\theta$, $\rho$ and $\lambda$ vs. Time $t$ in Gyr.

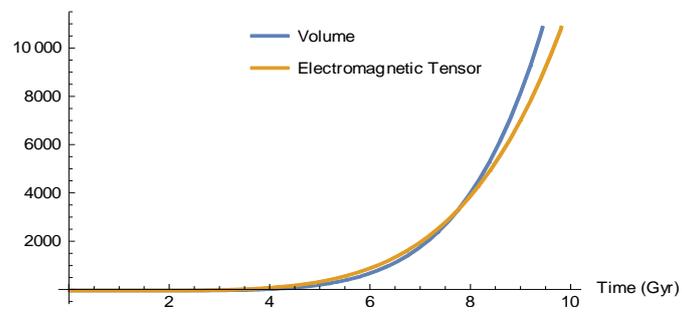

Figure 3: Variation of $V$, $F_{14}$ vs. Time $t$ in Gyr.

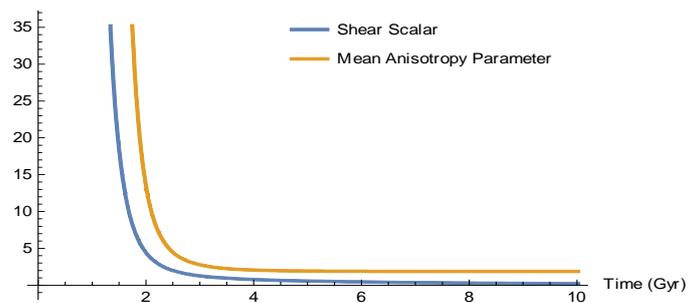

Figure 4: Variation of $\sigma$, $\Delta$ vs. Time $t$ in Gyr.

(i) The expression (24) shows that the deceleration parameter $q$ is constant and negative. This represents that our proposed model universe has been expanding with constant acceleration since cosmic time $t = 0$. The expansion rate of the universe is uniform throughout the time of evolution. For $m = 0.5$, $q=$ constant $= -0.5$, as shown in fig 1. Also, it is seen that the particle density has a large negative value at the time of the big bang when $t = 0$ as shown in fig 1. As time passes, it approaches positive value and reaches to a constant finite value at $t \to \infty$. This shows that the particle dominates the universe as time passes and corresponds to total constant finite number of particles in the universe.

(ii) The Hubble's parameter, expansion scalar, energy density, tension density show similar variations with time as shown in fig. 2. Both the Hubble's parameter $H$ and expansion scalar $\theta$ decreases with cosmic time $t$ after evolving from infinity at $t = 0$ and reaches to a finite value at $t \to \infty$. Also, we found that $\frac{dH}{dt}$ is negative which suggest that our universe is expanding with an accelerated rate. From the expression (31), it is observed that the rest energy density $\rho$ is a decreasing function of cosmic time $t$. It is seen that the rest energy density always remains positive $\rho \geq 0$ for all $m \geq -\frac{1}{2}$. The reason for getting positive energy density is that as we have known that the universe is expanding, this expansion is tending to speed up. Initially when $t \to 0$, then $\rho \to \infty$, and thus has an initial singularity. The particle density has also initial singularity at $t = 0$ and decreases as a function of time and disappears. This suggests that the string dominates at the beginning of the universe and plays an important role in forming the structures like galaxies and cluster of galaxies. Later, they lose their energy by gravitational radiation and therefore shrink and disappear representing matter dominated at the present day universe.

(iii) From the expression (35), we observed that the proper volume $V$ is zero at the time of big-bang and is found increasing exponentially with cosmic time $t$. This means that the universe first starts with zero volume at $t = 0$ and later expands as a function of cosmic time. At $t \to \infty$, the proper volume $V$ becomes infinite. The non-vanishing electromagnetic field tensor $F_{14}$ given by equation (33) shows that $F_{14}$ increases exponentially as a function of cosmic time $t$ as depicted in fig 3. We see that the electromagnetic field tensor $F_{14}$ does not vanish if $a \neq 0$. It has considerable influence in building up strings at early stage of evolution of the universe. The string and electromagnetic field are found to co-exist together in this model. The behaviour of proper volume and electromagnetic field tensor with time is shown in fig 3.

(iv) The parameters shear scalar $\sigma^2$ and mean anisotropy parameter $\Delta$ diverge at the initial singularity as demonstrated in fig 4. The model represents a shearing and non-rotating universe which shows a possible big crunch at some initial epoch $t = 0$. Also, from the mathematical the expression (40), we see that $\lim_{t \to \infty} \frac{\sigma^2}{\theta^2} \neq 0$ (constant), so, the model is an anisotropic one. Though anisotropy is induced to the system, it makes no contradiction to the present day's observation which tells that our universe posses isotropy. This is because during the process of evolution, the initial anisotropy disappears after some duration and approaches to isotropy and eventually it evolves into a FRW universe as suggested by [39].

In absence of magnetic field, we obtain the following field equations with the same assumption we used with presence of magnetic field

$$\frac{\ddot{B}}{B} + \frac{\ddot{C}}{C} + \frac{\dot{B}\dot{C}}{BC} = \lambda \tag{41}$$

$$\frac{\ddot{A}}{A} + \frac{\ddot{C}}{C} + \frac{\dot{A}\dot{C}}{AC} = 0 \tag{42}$$

$$\frac{\ddot{A}}{A} + \frac{\ddot{B}}{B} + \frac{\dot{A}\dot{B}}{AB} = 0 \tag{43}$$

$$\frac{\dot{A}\dot{B}}{AB} + \frac{\dot{B}\dot{C}}{BC} + \frac{\dot{A}\dot{C}}{AC} = \rho \tag{44}$$

Following the same techniques with presence of magnetic field, in this case also, the expressions for various physical parameters can be found out.

## 6.  Concluding Remarks

In the general theory of relativity, the interaction of electromagnetic field with an anisotropic Bianchi type-I string cosmological model universe is studied. Here, in this model $H > 0$, and $q < 0$. This shows that the universe is expanding exponentially which start with big-bang at the cosmic time $t = 0$ with zero volume and expands with an acceleration. The relation between Hubble's parameter and average scale factor as assumed in (21) yields a constant negative value of deceleration parameter for our model universe. The derived model has a point type [40] singularity at $t = 0$. Our model satisfies the energy conditions $\rho \geq 0$ and $\rho_p \geq 0$. The electromagnetic field affect the behaviour of the model and the expressions for physical parameters get modified. In presence of electromagnetic field, we see that the rest energy density and string tension tensity decreases as a function of cosmic time $t$. Also, the particle density and string tension density are comparable whereas the string tension density vanishes more rapidly than the particle density. This shows that at late time stage our model represents a matter dominated universe, that the present day's observational data. In our model, the universe may have some probability to be anisotropy throughout the evolution from early to late time stage. Recent observations tells that there is a discrepancy in measuring intensities of microwaves coming from different directions of the sky. This put forward us to study the universe with anisotropic Bianchi type I metric in such a way to describe our universe in more realistic situation. Also, several CMB anomalies such as inconsistency of temperature anisotropies in the CMB with exact homogeneous and isotropic FRW model measured by COBE/WMAP satellites, foregrounds/systematics and exotic topologies are an evidence that we live in a globally anisotropic universe. During inflation, shear decreases and eventually it turns into an isotropic phase with negligible shear. So in recent times, one needs to induce anisotropy in space-time in order to produce any substantial amount of shear.

**Acknowledgments**

The authors are thankful to CSIR, India for providing fund under the sanction **CSIR-HRDG Scheme No. 25(0279)/18/EMR-II** of dated **4th April, 2018**, to carry out this work successfully. Also, the authors are very thankful to the honorable referees for their valuable comments and suggestions for further improvement of the paper.